# Mechanism of photo-excited precession of magnetization in (Ga,Mn)As on the basis of time-resolved spectroscopy

T. Matsuda* and H. Munekata

Imaging Science and Engineering Laboratory, Tokyo Institute of Technology, Japan

4259-J3-15 Nagatsuta, Midori-ku, Yokohama 226-8503, Japan

**Abstract**

Aiming at studying the mechanism of photo-excited precession of magnetization in ferromagnetic $Ga_{1-x}Mn_xAs$, magneto-optical (MO) and differential reflectivity ($\Delta R/R$; DR) temporal profiles are studied at relatively long (ps − ns) and ultra-short (1 ps or less) time scales for samples with different Mn contents ($x = 0.01 - 0.11$). As to the oscillatory MO profiles observed in the long time scale, simulation based on the Landau-Lifshits-Gilbert equation combined with two different magneto-optical (MO) effects confirms photo-inducement of the perpendicular anisotropy component $\Delta H_{\text{eff}, \perp}$. As for the profiles observed in the ultra-short time scale, they are consistently explained in terms of the dynamics of photo-generated carriers, but not by the sudden reduction in magnetization (the ultrafast demagnetization). With those experimental results and analyses, new mechanism which accounts for the photo-induced $\Delta H_{\text{eff},\perp}$ is addressed: namely, photo-ionization-like excitation of $Mn^{2+}$, $Mn^{2+} + h\nu \rightarrow Mn^{2+,*} = Mn^{3+} + e^-$. It is discussed that such excitation tips magnetic anisotropy toward the out-of-plane direction through the inducement of orbital angular momentum and the gradient $\partial[Mn^{2+,*}]/\partial z$. Validity of the proposed mechanism is examined by estimating the efficiency of excitation on the basis of the Lambert-Beer law and the experimental $\Delta H_{\text{eff},\perp}$ values, through which the efficiency of 1 - 10 ppm with the nominal optical cross section of around $5 \times 10^{-12}$ m$^2$ are obtained.

## I. INTRODUCTION

The present work concerns with photo-excited precession of magnetization (PEPM) in the *p*-type, ferromagnetic semiconductor (Ga,Mn)As [1-10]. It has been established that PEPM in *p*-(Ga,Mn)As is triggered by femto-second (fs) laser pulses of photon energy near the GaAs band gap with relatively low laser fluence (0.1 − 10 μJ/cm$^2$), without external fields or angular momentum of light. This fact indicates that the weak excitation itself tips the effective field $H_{eff}$ off from the equilibrium and exerts torque on magnetization *M*, which exceedingly contrasts with PEPM in metals and insulators [11, 12].

Researchers have inferred that direction of the $H_{eff}$ vector varies in the plane as the consequence of imbalance between bulk-cubic and interface-uniaxial anisotropies due to ultrafast rise in either hole concentration or temperature of spin sub-system [2-4, 7-9]. This scenario, however, has been opposed by the work which analyzes both oscillatory and dc-components of PEPM with simulation based on the Landau-Lifshits-Gilbert equation combined with two different magneto-optical (MO) effects, polar Kerr rotation $I_z$ and magnetic birefringence $I_{xy}$ [13, 14] (the LLG-2MO simulation, Fig. 1): the $H_{eff}(t)$ vector, whose blunt dynamic response differs from abrupt optical response induced by photo-generated carries [15], is found to be tipped toward out-of-plane [5,6]. While the photo-induced perpendicular anisotropy component, $\Delta H_{\text{eff}, \perp}$, has been confirmed by others [9], the mechanism of $\Delta H_{\text{eff}, \perp}$ remains open and needs to be explained by the model beyond the *p-d* Zener model [16, 17] which only assumes the interaction between $Mn^{2+}$ ions and free holes.

Motivated by the above-stated background, optical responses of various $Ga_{1-x}Mn_xAs$ samples with different Mn contents *x* have been investigate at two different time scales; the relatively long time scale at which precession of magnetization takes place (ps to ns) and so-called ultra-short time scale (1 ps or less) at which precursory process which triggers PEPM may be optically detected. Temporal MO profiles in the long time scale have been successfully modeled by the LLG-2MO simulation with only introducing photo-induced perpendicular component, $\Delta H_{\text{eff}, \perp}$. The simulation reveals that $\Delta H_{\text{eff}, \perp}$ is generated relatively faster (10 − 20 ps) for below-gap excitation than for above-gap excitation (~ 100 ps), suggesting that the pathway of energy transfer from electrons to spins is dependent on the band structure near the band edges. In the ultra-short time scale, rapidly oscillating signals have been observed for below-gap excitations whereas spike-like signals have appeared for above-gap excitation (except for the $x$ = 0.01 sample). The autocorrelation approach has been adopted to analyze those signals, which suggests that both types of signals reflect dynamics of photo-generated carriers but not directly associated with spin/magnetization dynamics. Simulation with the Landau-Lifshitz-Bloch (LLB) equation including longitudinal relaxation of macro-spins has failed to reproduce the experimental *positive* spikes in sense that the LLB simulation always yields *negative* spikes as the consequence of ultrafast reduction of in-plane magnetization.

On the basis of those findings, we have concluded that ultrafast demagnetization is not the primary mechanism which triggers PEPM, and have addressed new mechanism for the photo-induced $\Delta H_{\text{eff}, \perp}$: that is photo-ionization-like excitation of $Mn^{2+}$, $Mn^{2+}$ + $h\nu \rightarrow Mn^{2+,*} = Mn^{3+} + e^-$. Here, what brings about the anisotropy along the $z$ axis is orbital angular momentum of the $Mn^{3+}$ component [18] together with the dynamic concentration gradient $\partial[Mn^{2+,*}]/\partial z$. Efficiency of excitation has been estimated to be 1 - 10 ppm together with the nominal optical cross section of around $5 \times 10^{-12}$ $m^2$. The relatively large value of cross section could be regarded as a clue to quantify the magnitude of *p-d* hybridization around the valence band top discussed theoretically [19, 20].

## II. EXPERIMENT

Four, 100 nm-thick $Ga_{1-x}Mn_xAs$ epi-layers ($x$ = 0.01, 0.02, 0.08, 0.11), grown on LT-GaAs/GaAs(001) substrates at 235 °C by molecular beam epitaxy, were studied. They show in-plane magnetic easy axis nearly along GaAs ⟨100⟩ axis at low temperatures (< 20 K), and exhibit so-called metallic conduction at low temperatures [21]. Temporal profiles of magneto-optical (MO) and differential reflectivity ($\Delta R/R$; DR) signals were measured by one-color pump-and-probe (P-P) system based on a mode-locked Ti:sapphire laser whose pulse duration and repetition rate were around 150 fs, and 76 MHz, respectively. Wavelength of the laser was varied between 750 and 900 nm ($h\nu$ = 1.38 to 1.65 eV). Experimental setups for the MO and DR measurements are shown schematically in Figs. 2(a) and (b), respectively.

As summarized schematically in Fig. 1, the LLG-2MO simulation handles both oscillatory and dc-components of MO temporal profiles, and extracts dynamics of photo-induced $H_{\text{eff}}(t)$. Because of this reason, great care was taken to keep a stable optical base line extended up to 3 ns as well as high signal-to-noise ratio: positions of a collimator and a retroreflector in an optical delay line were precisely adjusted so as to retain the same beam diameter at the point 10 m far from the delay line throughout the entire range of the delay time (Fig. 2). Pump and probe beams, whose polarizations are both *E*//GaAs [010], were focused into the same spot of around 100 μm in diameter using a CCD camera equipped with a long-focus microscope. Precision of the overlap was 10 - 20 μm. The incident angles with respect to the axis normal were around 6 and 3° for pump and probe beams, respectively. A reflected probe beam was guided into an optical bridge which was placed 1.8 m far from the sample plane. Time interval of probing in the ultrafast time region (t < 4 ps) was 26 fs as determined by the mechanical precision of a mechanical delay stage. All P-P measurements were carried out at 10 K. Prior to the measurements, samples were magnetized by applying an external field of *B* = 0.2 Tesla along the in-plane, [010] direction. No external fields were applied during the measurements.

## III. RESULTS

Shown in Fig. 3(a) are long-time-scale temporal profiles of MO signals obtained for the $x = 0.02$ sample at six different P-P photon energies. No background subtraction was carried out from the raw MO data. A resistance-temperature curve of the $x = 0.02$ sample is depict in the inset Fig. 3(a). The fluence of pump and probe pulses were kept constant at $I_{pump} = 1.7$ $\mu$J/cm$^2$ and $I_{probe} = 84$ nJ/cm$^2$, respectively. In the MO profile taken at $h\nu = 1.57$ eV, note the presence of a spike-like component in the time domain of a few ps, as highlighted by a rectangle, in addition with damping oscillation due to PEPM. This component always appears on the same side of the first peak of oscillations. At $h\nu = 1.51$ eV, being nearly the band-edge excitation, amplitude of the oscillation and height of the spike both are reduced. Simultaneously, phase of the oscillatory component shifts toward shorter time scale. A large, long-lived exponential component with the lifetime of around 2300 ps is presumed to be carrier dynamics in a GaAs substrate [15, 22]. Further decreasing P-P photon energy ($h\nu \leq 1.48$ eV) results in disappearance of the spike-like component, whereas damping oscillation remains in the profiles down to $h\nu = 1.41$ eV ($\lambda = 880$ nm) without any further phase shift. The observed, gradual reduction in PEPM amplitude with decreasing photon energy suggests the excitation of the band-edge tails caused by Mn-related local potential fluctuations in (Ga,Mn)As [23].

Graphical representations of fitting experimental data with the LLG-2MO simulation are shown in Fig. 3(b). For both above- and below-gap excitations, the entire PEPM profile is well reproduced by the simulation assuming the change in $H_{\text{eff}}$ toward the z axis with the rotation function $\theta(t) = \theta_0\{1-\exp(-t/\tau_1)\}\exp(-t/\tau_2)$ in which $(\tau_1, \tau_2)$ = (90, 120) and (20, 200) in the unit of ps for the excitation at $h\nu$ = 1.57 and 1.44 eV, respectively. Physical quantities associated with PEPM are summarized in Table 1 for different photon energies. The relatively small $\tau_1$ value (fast rising of the $\Delta H_{\text{eff},\perp}$) at the excitation $h\nu$ = 1.4 ~ 1.5 eV suggests that below-gap photons excite the states which couples more efficiently with ferromagnetically ordered spins. Study in the range $h\nu \geq$ 1.57 eV reveals slower rising of $\Delta H_{\text{eff},\perp}$ with increasing $h\nu$ values, which we discuss in a separate paper in connection with time dependent MO coefficients induced by photo-generated free carrier [24, 25].

Polarization dependences of pump/probe pulses on temporal MO profiles are shown in Fig. 4(a) for below-gap excitation. Similar to those obtained by above-gap excitation [5], inversion of oscillation phase is clearly seen when polarization of probe pulses is rotated 90°, in between [010] and [100]. At the intermediate polarization of [110], oscillation is suppressed. The observed results indicate that precession of magnetization is detected primarily through magnetic birefringence (MB) term $I_{xy}$ [13, 14]. Negligibly small contribution of the polar Kerr rotation (PKR) term $I_z$ at $h\nu < 1.51$ eV is most likely due to multi-layer interference [14] and wavelength-dependent complex refractive index [26]. See Table 1 for the complicated $x$ dependence on the ratio of MO coefficients. It is worth stressing that, if ***M*** starts precessing toward out-of-plane direction ($z$ axis), the amplitude of oscillation would not be as large as the experimental data (Fig. 4(a)) for the first tens of ps due to the small $I_z$ value. Besides, the phase of oscillation would not change with polarization of probe pulses, if PKR is the primary MO effect which yields the observed MO signals.

We now turn our eye on the time scale less than a few ps. Shown in Fig. 5(a) are MO temporal profiles obtained at the excitation energy of $h\nu = 1.44$ eV for various pump fluences $I_{pump}$ ranging between 0.34 and 10 $\mu$J/cm$^2$. We find rapidly oscillating signals around the time zero throughout the entire range of $I_{pump}$. Temporal DR profile also exhibits similar kind of oscillating signals (Fig. 5(b)), indicating that the observed rapid oscillation is attributed to the change in refractive index induced by photo-excited electronic polarization **P** [27].

Analysis of rapid oscillation has been carried out on the basis of autocorrelation function [28], which is represented by Eqs.1a and b.

$$\Delta r(t) = \Delta r^0 \int_{-\infty}^{t} f(t-\tau) \left| E_{pu}(\tau) + E_{pr}(\tau) \right|^2 d\tau \qquad (1a)$$

$$S^{(1)}(t) = \int_{-\infty}^{\infty} \left| \{r^0 + \Delta r(\tau')\} E_{pr}(\tau') \right|^2 d\tau' \qquad (1b)$$

Eq. 1a represents optical response at a sample surface $\Delta r(t)$ induced by electric fields of pump and probe pulses, $E_{pu}(t) = E_{pu}^0(t)\cdot\exp(-i\omega t)$ and $E_{pr}(t) = E_{pr}^0(t)\cdot\exp(-i\omega t)$, respectively, whereas Eq. 1b the intensity of reflected probe light $S^{(1)}(t)$ affected by the coherent component of a reflected light pulse. Pulse envelopes are expressed by $I_{pu}(t) = I_{pu}^0 G(t) = |E_{pu}^0(t)|^2$ and $I_{pr}(t - t_{delay}) = I_{pr}^0 G(t - t_{delay}) = |E_{pr}^0(t - t_{delay})|^2$ in which $t_{delay}$ is the time delay of a probe pulse and $G(t)$ the Gaussian function. The calculated profile which reproduces the experimental data (Fig. 5(b)) is obtained with $\Delta r^0/r^0 = 0.01$ and the response function $f(t) = \exp(-t/40 \text{ fs})$, together with the experimental conditions of $I_{pu}^0$

= 1.7 $\mu$J/cm$^2$, $I_{pr}^0$ = 84 nJ/cm$^2$, and the standard deviation in G($t$), $\sigma$ = 38.6 fs. The fast de-coherence time constant of $T$ = 40 fs in $f(t)$ can not be attributed to spin-flip scattering since ferromagnetic order persists during the excitation. Presence of other mechanism was suggested in the studies of four-wave mixing (FWM) spectroscopy in CdMnTe under high magnetic fields [29] and (Ga,Mn)As with $x \leq 0.001$ [30], but was left open. We discuss this point later in relation with photo-ionization like excitation of Mn$^{2+}$.

Phase inversion by rotating 90° the polarization of probe pulses is not clearly established in rapidly oscillating MO signals (Fig. 4(b)). Influence of pump pulse polarization on oscillation phase is rather suggested in the signals. In view of further digging for spin dynamics in ultra-short time scale, it is interesting to examine experimental data with more precise simulation which takes into account ultrafast change in the polarization of $E_{pu}$ caused by both carrier and spin dynamics. Further precise measurements together with such advanced simulation would shed light on disturbance of ordered spins in the ultrafast time scale.

A rapidly oscillating component is replaced by a spike-like component for the above-gap excitation, as shown in Fig. 5(c). The spike-like component shown in Fig. 3(a) is thus reconfirmed by this measurement. Coexistence of the oscillatory component is also noticeable at relatively high $I_{\text{pump}}$. A DR temporal profile also shows mixture of those two components (Fig. 5(d)), which indicates that the observed spike is again attributed to the change in refractive index but now accompanied by the process which destroys in part the phase coherence between $E_{pu}(t)$ and $E_{pr}(t)$, namely, by the thermalization of photo-generated free carriers [27]. With this inference in mind, Eq. 1b is modified into Eq. 2 by omitting the interference terms $E_{pu}^{*}E_{pr}$ and $E_{pu}E_{pr}^{*}$.

$$S^{(2)}(t) = \int_{-\infty}^{\infty}\{r^0 + \Delta r(\tau')\}I_{pr}(\tau')d\tau' \qquad (2)$$

Calculated profiles using Eq. 2 are shown by closed symbols in Fig. 5(c) and a top-most solid line in Fig. 5(d). Through the calculations, we find that a response function consisting of the sum of two exponential decay components, $f(t) = \exp(-t/180\text{ fs}) + 0.12\exp(-t/900\text{ fs})$, reproduces the experimental data. Referring the values of the time constants obtained by the calculation, we infer that the first exponential term ($T = 180$ fs) is associated with LO phonon scattering [31, 32] whereas the second term ($T = 900$ fs) with carrier trapping [33] or electron-hole-pair scattered by the background free holes [34]. While analysis using the incoherent part of autocorrelation stays at the level of phenomenological interpretations, it is clear that the spike-like feature appearing in case of above-gap excitation can be explained consistently in terms of thermalization of photo-generated carriers.

In order to further examine the effect of ultrafast heating caused by stronger optical absorption above the band-gap energy, calculation of MO profiles due to hypothetical demagnetization has been carried out using the Landau-Lifshitz-Bloch (LLB) equation [35].

$$\frac{\partial M}{\partial t} = \gamma M \times H + \gamma M_s \frac{\alpha_\perp}{M^2} M \times (M \times H) - \gamma M_s \frac{\alpha_{//}}{M^2} (M \cdot H) M \qquad (3)$$

Here, $\gamma$ is the gyromagnetic constant, $M_s$ saturation magnetization, $H$ the effective magnetic field, whereas $\alpha_\perp \sim 0.25$ (Table 1 in the main text) and $\alpha_{//}$ are dimensionless transverse and longitudinal damping factors, respectively. Ultrafast demagnetization is expressed by the exponential function $\alpha_{//}(t) = \alpha_{//}^0 \exp(-t/\tau_{dem})$ with its magnitude $\alpha_{//}^0 =$ 0.09 and 0.15 for $h\nu =$ 1.44 and 1.57 eV, respectively. The lifetime $\tau_{dem} \sim$ 0.4 ps is chosen in view of reproducing the experimental data. $H$ is assumed $H = H_0 + H_{ext} + H_{dem}$ in which the in-plane crystal anisotropy field $H_0 \sim$ 2000 Oe, an external field $H_{ext}$

= 0, and the demagnetizing field $\boldsymbol{H}_{\text{dem}} = 0$.

A *negative* spike-like component is obtained, reflecting the reduction of in-plane magnetization ($M_x$). In detail, our calculation yields polarization rotation of 3.7 and 1.4 $\mu$rad, for $h\nu$ = 1.57 and 1.44 eV, as shown by open circles in Figs. 6(a) and (b). Those values correspond to a reduction in magnetization of 0.19 % (Fig. 5(d)) and 0.13 %, respectively, referring the value of MO rotation measured separately in the process of the 90 ° magnetization switching (Table 1). Consequently, we find that the LLB simulation always yields a *negative* spike which is opposite to the observed spikes. Therefore, we infer that the observed spike-like component is primarily attributed to the dynamics of photo-generated carriers. Otherwise, signals due to the ultrafast demagnetization are small so that they are masked behind the signals due to photo-generated carriers.

Interesting sample dependence is observed in the ultra-short time scale (Fig. 7(a)). Rapidly oscillating component dominates for both below-and above-gap excitations in the $x$ = 0.01 sample which is near the boundary between insulator and metallic conditions [21]. On the other hand, spike-like components are conspicuous in the $x$ =

0.1 sample which is strongly metallic.

The MO profiles in the long time scale obtained by below-gap excitation (Fig. 7(b)) show that precession frequency decreases with increasing the $x$ values, whereas damping tends to minimize at the intermediate $x$ values. The observed trends are qualitatively similar to those obtained by above-gap excitation [36]. Analytical results obtained by the LLG-2MO simulation are compiled in Table 1. We find that PEPM dynamics becomes less dependent on P-P photon energy for the $x = 0.08$ sample.

The $\tau_1$ value, which represents the generation rate of $\Delta H_{\text{eff}, \perp}$, is nearly constant at the P-P photon energy of $h\nu = 1.57$ eV for all three samples, whereas it becomes shorter when excited at $h\nu = 1.44$ eV, in particular, for the $x = 0.01$ and $0.02$ samples. On the other hand, the $\tau_2$ value, being the relaxation rate of $\Delta H_{\text{eff}, \perp}$, is obviously larger than the $\tau_1$ value and tends to scatter among different samples, suggesting that relaxation process is significantly influenced by extrinsic sources such as defects in the bulk and surfaces.

## IV. DISCUSSION

In (Ga,Mn)As, it is experimentally established that the upper part in the valence band is strongly hybridized with Mn $d$-orbital [37-39], as shown schematically in Fig. 8. This gives rise to the notion that the interband excitation involves two different transitions, hole-electron pair generation and photo-ionization-like transition, $Mn^{2+} + h\nu \rightarrow Mn^{3+} + e^-$. The change in magnetization caused by the latter transition can be expressed by $\delta M^* = M^*(\text{light}) - M(\text{dark}) = \Delta Mn^{2+,*} \cdot [p_m(Mn^{3+}) - p_m(Mn^{2+})]$. Here, $\Delta Mn^{2+,*}$ is the number of excited $Mn^{2+}$ ions [m$^{-3}$] in the unit sheet of (Ga,Mn)As with the thickness of one lattice constant ($a_0 = 0.565$ nm), whereas $p_m(Mn^{3+})$, and $p_m(Mn^{2+})$ are magnetic moments of $Mn^{3+}$ and $Mn^{2+}$, respectively. Assuming the $g$-factor $g = 2.7$ [18] and the net spin number $S = 4/2$ for $Mn^{3+}$, and $g = 2.0$ and $S = 5/2$ for $Mn^{2+}$, we obtain $\delta M^* = 0.4 \cdot \mu_B \cdot \gamma \cdot \hbar \cdot \Delta Mn^{2+,*} = 4.68 \times 10^{-30} \cdot \Delta Mn^{2+,*}$ with the unit of [N/A·m] or [Wb·m$^{-2}$]. Here, $\mu_B$, $\gamma$, and $\hbar$, are Bohr magneton, gyromagnetic ratio ($e/2m$), and Plank constant, respectively. The increment in the $\Delta M^*$ is attributed to the large $g$-factor of $Mn^{3+}$, the orbital angular momentum. Namely, it can conceptually be regarded as the magnetization generated by a virtual current (a flow of kinetic energy) passing through a virtual solenoid (orbital) whose central axis is parallel to the $z$ axis. Naturally, one of the most likely sources which yield such a current is the electrochemical potential gradient $\partial Mn^{2+,*}/\partial z$ (inset Fig. 8).

Since the photo-induced effective field per sheet is expressed by $\delta H_{eff,\perp} = \delta M^*/\mu_0 = 3.71 \times 10^{-24} \cdot \Delta Mn^{2+,*}$ [A/m] with $\mu_0$ the vacuum permeability, the overall field induced by the photo-excitation is obtained by integrating $\delta H_{eff,\perp}$ over the entire layer thickness $d$ in the unit of the lattice constant $a_0$ ($d = 100$ [nm] $= 1.77 \times 10^{11}$ [$a_0$]). On the basis of Lambert-Beer law, the number of $Mn^{2+,*}$ produced in the unit sheet at the depth $z$ (in the unit of the lattice constant $a_0$) is expressed by $\Delta Mn^{2+,*}(z) = \beta \cdot [Mn^{2+}] \cdot \alpha \cdot I = \beta \cdot [Mn^{2+}] \cdot \alpha \cdot I_0 \exp(-\alpha \cdot z)$ in which $\beta$, $[Mn^{2+}]$, $I_0$, and $\alpha$ are *nominal* absorption cross section [m$^2$], the number of $Mn^{2+}$ ions in the unit sheet, photon flux [m$^{-2}$], and absorption coefficient in the unit of inverse lattice constant [$a_0^{-1}$], respectively. Putting all together, we obtain Eqs. 4a and b for the overall photo-induced field $\Delta H_{eff,\perp}$ and the total number of photo-excited $Mn^{2+}$ ions, $[Mn^{2+,*}]$, respectively.

$$\Delta H_{eff,\perp} = \int_0^d \delta H_{eff,\perp}\, dz = 3.71 \times 10^{-24} \int_0^d \beta \cdot \Delta Mn^{2+,*} \cdot I_0 \exp(-\alpha \cdot z)\, dz \tag{4a}$$
$$= 9.30 \times 10^{-7}\, \beta \cdot I_0 \cdot \{1 - \exp(-\alpha \cdot d)\}$$

$$[Mn^{+2,*}] = 2.51 \times 10^{17} \cdot \beta \cdot I_0 \cdot \{1 - \exp(-\alpha \cdot d)\} \tag{4b}$$

The values $I_0\{1 - \exp(-\alpha \cdot d)\}$ on the RHS of Eqs. 4a and 4b are $7.35 \times 10^{14}$ and $6.4 \times 10^{15}$ photons/m$^2$, with $\alpha = 5.65 \times 10^{-5}$ [$a_0^{-1}$] ($= 10^3$ [cm$^{-1}$]), $I_0 = 7.4 \times 10^{16}$ photons/m$^2$, and $\alpha = 5.65 \times 10^{-4}$ $[a_0^{-1}]$ ($= 10^4$ cm$^{-1}$), $I_0 = 6.8 \times 10^{16}$ photons/m$^2$ at $h\nu = 1.44$ and 1.57 eV, respectively. As to the LHS of Eq. 4a, the peak $\theta_{\text{eff}}$ values of the temporal $\theta_{\text{eff}}$ profiles are obtained from the dotted lines in Fig. 3b, which are $\theta_{\text{eff, peak}} = 0.020$ and 0.21 μrad for $h\nu = 1.44$ and 1.57 eV, respectively. Then, the maximum $\Delta H_{\text{eff}, \perp}$ values are estimated to be $4.0 \times 10^{-5}$ and $3.8 \times 10^{-4}$ Oe for $h\nu = 1.44$ and 1.57 eV, respectively, using the $z$ component of the magnetization rotation function, $H_0 \cdot \sin\theta(t)$ (Fig. 1). We finally obtain $\beta = 4.7 \times 10^{-12}$ and $5.1 \times 10^{-12}$ m$^2$ for $h\nu = 1.44$ and 1.57 eV, respectively. Although it is not the rigorous comparison, yet these values are obviously larger than the optical cross section standards of transition metal and rare earth ions in optical materials including zincblende semiconductors [40-42]. Hybridization between Mn $d$ orbits and host valence bands may be responsible for the large $\beta$ values. The activation efficiencies [Mn$^{2+,}$*] / [Mn$^{2+}$] are estimated by using Eq. 4b, which yields $1.9 \times 10^{-6}$ and $1.8 \times 10^{-5}$ for $h\nu = 1.44$ and 1.57 eV, respectively.

The time lag ($\tau_1$) between the time of excitation and the time of $\Delta H_{\text{eff}, \perp}$ maximum suggests that the photo-ionization-like transition would be a subordinate process that follows the hole-electron pair generation. It is inferred that this transition occurs when excitation energy is transferred by the annihilation of hole-electron pairs (Auger recombination) or by the scattering between hole-electron pairs and $Mn^{2+}$ ions. Reduced $\tau_1$ value with decreasing excitation photon energy suggests that valence band states near the Fermi level can influence Mn spins relatively faster, which is in accordance with the picture of strong *p-d* hybridization near the valence band top [19, 20].

## V. CONCLUSIONS

We have presented time-resolved magneto-optical (MO) and differential reflectivity (ΔR/R; DR) measurements at relatively long (ps – ns) and ultra-short (1 ps or less ) time scales for $Ga_{1-x}Mn_xAs$ samples with different Mn contents ($x$ = 0.01 – 0.11). Inducement of perpendicular $H_{\text{eff}}$ component $\Delta H_{\text{eff},\perp}$ has been confirmed by the MO temporal profiles obtained at relatively long time scale for different excitation photon energies combined with various polarization combinations of probe pulses, and by the analysis of those profiles using the simulation based on Landau-Lifshits-Gilbert equation incorporating two different magneto-optical effects (the LLG-2MO simulation). It has also been found that $\Delta H_{\text{eff},\perp}$ is generated relatively faster (10 – 20 ps) for below-gap excitation than for above-gap excitation (~ 100 ps), suggesting that electronic states near the valence band top as well as those slightly inside the band gap couple efficiently with ferromagnetically ordered spins.

In the ultra-short time scale, rapidly oscillating and positive spike-like signals have been observed with below- and above-gap excitations, respectively. Those signals have been carefully analyzed by the autocorrelation approach, and have been concluded that they are most likely attributed to the dynamics of photo-generated carriers; otherwise signals due to ultrafast demagnetization is small enough to be masked behind the signals from those carriers. The fact that rapidly oscillating signals are not accompanied by the sudden, negative change in the base line at the time of excitation, the fingerprint of ultrafast demagnetization, indicate that, at least in the case of below-gap excitation, ultrafast demagnetization is not a requirement for the photo-induced $\Delta H_{\text{eff},\perp}$.

Motivated by all those experimental data, new mechanism of photo-induced perpendicular effective field $\Delta H_{\text{eff},\perp}$ has been addressed: photo-ionization-like excitation of $Mn^{2+}$, $Mn^{2+} + h\nu \rightarrow Mn^{2+,*} = Mn^{3+} + e^-$. Here, $Mn^{3+}$ represents the spin configuration of $3d^4$ with $g = 2.7$, whereas $e^-$ an electron in the conduction band or its tail states. It has been discussed that such excitation alters magnetic anisotropy toward the out-of-plane direction ($z$ axis) through the inducement of orbital angular momentum and the gradient $\partial[\text{Mn}^{2+,*}]/\partial z$. The validity of the proposed mechanism has been examined through the estimation of the nominal optical cross section $\beta$ on the basis of the Lambert-Beer law and the peak values in the dynamic $\Delta H_{\text{eff},\perp}(t)$ curves obtained by experiments. The $\beta$ value of around $5 \times 10^{-12}$ m$^2$ has been obtained, which would reflect strong $p$-$d$ hybridization around the valence band top.

We thank K. Nishibayashi and N. Nishizawa for their technical advices on optical experiments and samples, and acknowledge partial supports from Advanced Photon Science Alliance Project from MEXT and Grant-in-Aid for Scientific Research (No. 22226002) from JSPS.

**References**

* Present address: Advanced Technology R&D Center, Mitsubishi Electric Corp.,

8-1-1 Tsukaguchi-Honmachi, Amagasaki, Hyogo 661-8661, Japan.

2005), p. 342.

| Mn cont. | $T_C$ | $h\nu$ | $H_{\text{eff}}$ | $\alpha$ | $\theta_0$ | $\tau_1$ | $\tau_2$ | $I_z / I_{xy}$ |
|---|---|---|---|---|---|---|---|---|
| $x$ | (K) | (eV) | (Oe) | — | (μdeg) | (ps) | (ps) | — |
| 0.01 | 45 | 1.57 | 2600 | 0.27 | — | 80 | 2000 | − 10 |
| | | 1.44 | 2800 | 0.40 | — | 10 | 450 | − 15 |
| 0.02 | 47 | 1.57 | 1800 | 0.25 | 38 | 90 | 120 | + 1.5 |
| | | 1.48 | 2000 | 0.25 | 16 | 20 | 200 | 0.0 |
| | | 1.44 | 2000 | 0.25 | 8 | 20 | 200 | 0.0 |
| | | 1.41 | 2000 | 0.25 | 6 | 20 | 200 | 0.0 |
| 0.08 | 110 | 1.57 | 1055 | 0.06 | — | 90 | 1400 | − 2.7 |
| | | 1.44 | 1055 | 0.08 | — | 90 | 1400 | − 3.1 |

Table 1: Parameters of PEPM dynamics extracted by the LLG-2MO simulation using temporal MO profiles obtained from three samples with different Mn contents. Rotation angle $\theta_0$ of the $x = 0.02$ sample was determined by magneto-optical hysteresis data measured independently.

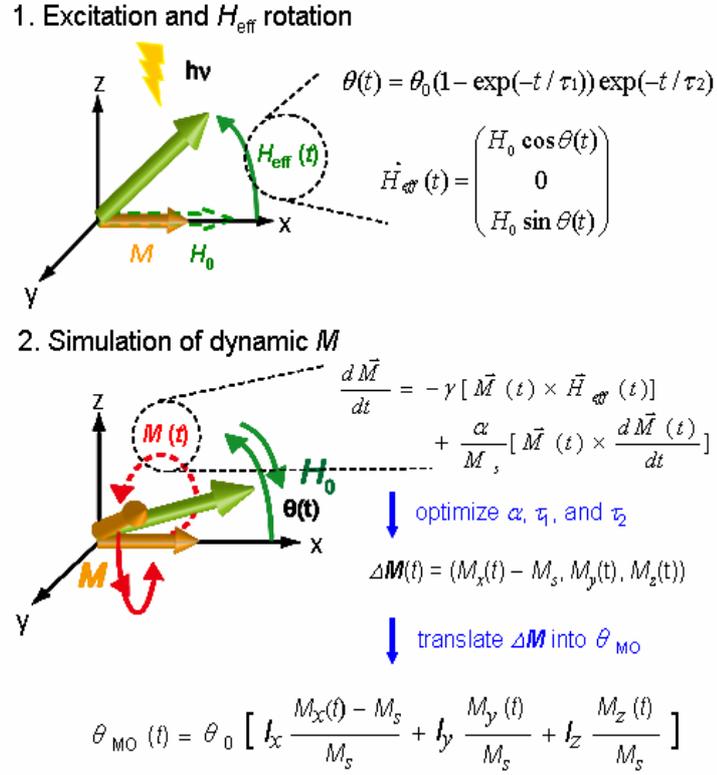

Figure 1: Schematic illustration of steps in the LLG-2MO simulation. A laser pulse results in a change in the direction of an effective magnetic field $\boldsymbol{H}_{\text{eff}}(t) = (H_0 \cos\theta(t), 0, H_0 \sin\theta(t))$ with dynamic rotation function $\theta(t)$. A magnetization vector $\boldsymbol{M}(t)$ precesses around the $\boldsymbol{H}_{\text{eff}}(t)$ with natural damping, and yields differential MO signals $\Theta_{\text{MO}}(t)$ which is the linear combination of magnetic birefringence $I_x \cdot \Delta M_x$ and $I_y \cdot \Delta M_y$, and polar Kerr rotation $I_z \cdot \Delta M_z$. $\alpha$ and $\gamma$ in the LLG equation are the effective Gilbert damping coefficient and gyromagnetic constant, respectively.

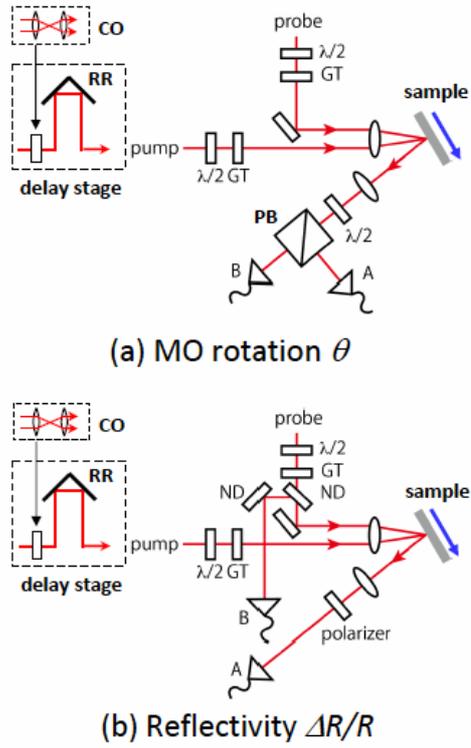

Figure 2: Schematic illustrations of experimental setups for (a) MO and (b) DR measurements. Delay stages built at the upstream of the pump lines are also shown in the insets. RR, λ/2, GT, PB, ND, CO represent, respectively, retroreflector, half-wave plate, Glan-Thompson prism, polarizing beamsplitter, neutral density filter, and collimator. CO consists of a pair of lenses. Optical bridge is composed of photo-diode detectors A and B. MO rotation $\theta$ is obtained by $\theta = (A-B)/2B$, $A \approx B$, whereas differential reflectivity by $\Delta R/R = (A-B)/2B$. Arrows behind samples represent in-plane ($x$-$y$ plane), remnant magnetization of samples. Photo-induced anisotropy occurs along the out-of-plane, $z$ axis, as shown in Fig. 1.

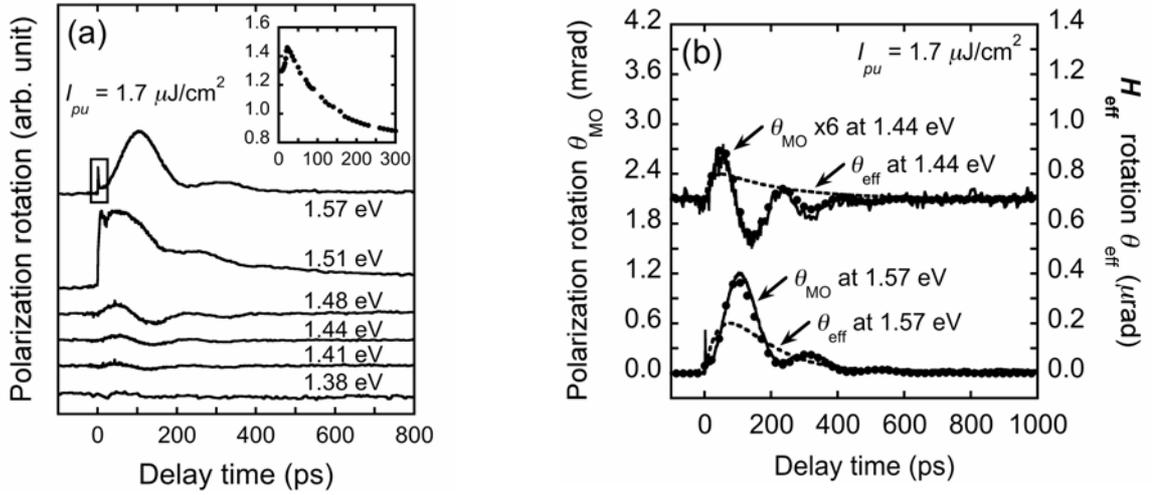

Figure 3: (a) Temporal MO profiles obtained from the $x = 0.02$ sample with six different P-P photon energies. Pump fluence is fixed at $I_{pump} = 1.7$ µJ/cm$^2$. Inset shows temperature dependence of sample resistance; horizontal and vertical axes represent temperature (K) and resistance (kΩ), respectively. (b) Simulated MO profiles (dots) and experimental MO profiles (solid lines) for above- and below-gap excitation, together with profiles of dynamic rotation function $\theta_{eff}$ (dashed lines).

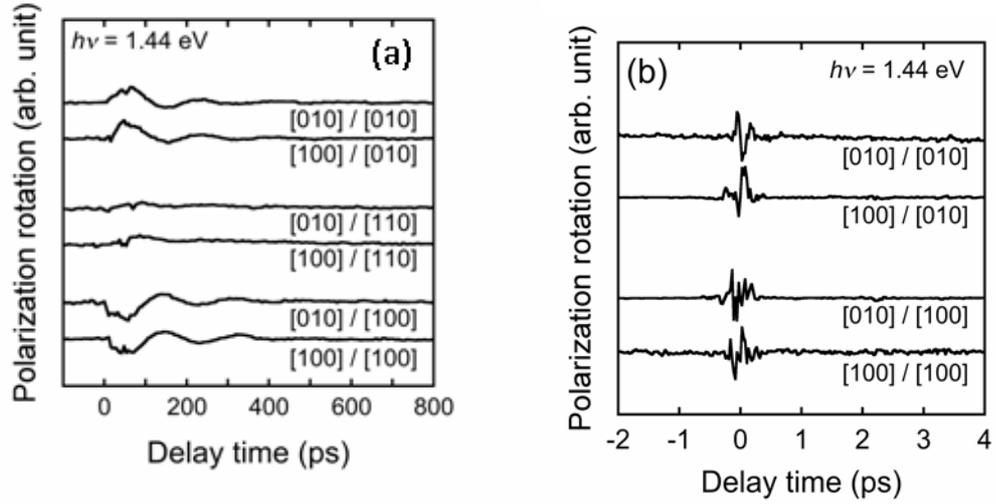

Figure 4: Temporal MO profiles obtained with various combinations of pump and probe polarization for (a) long and (b) short time scales. P-P photon energy is $h\nu = 1.44$ eV. The label [xxx] / [xxx] below each profile specifies the polarization of pump and probe pulses, respectively; for instance, [100] / [010] represents pump polarization of [100] and probe polarization of [010]. All experimental data were obtained at the pump and probe fluences of $I_{pump} = 1.7$ μJ/cm$^2$ and $I_{probe} = 84$ nJ/cm$^2$, respectively.

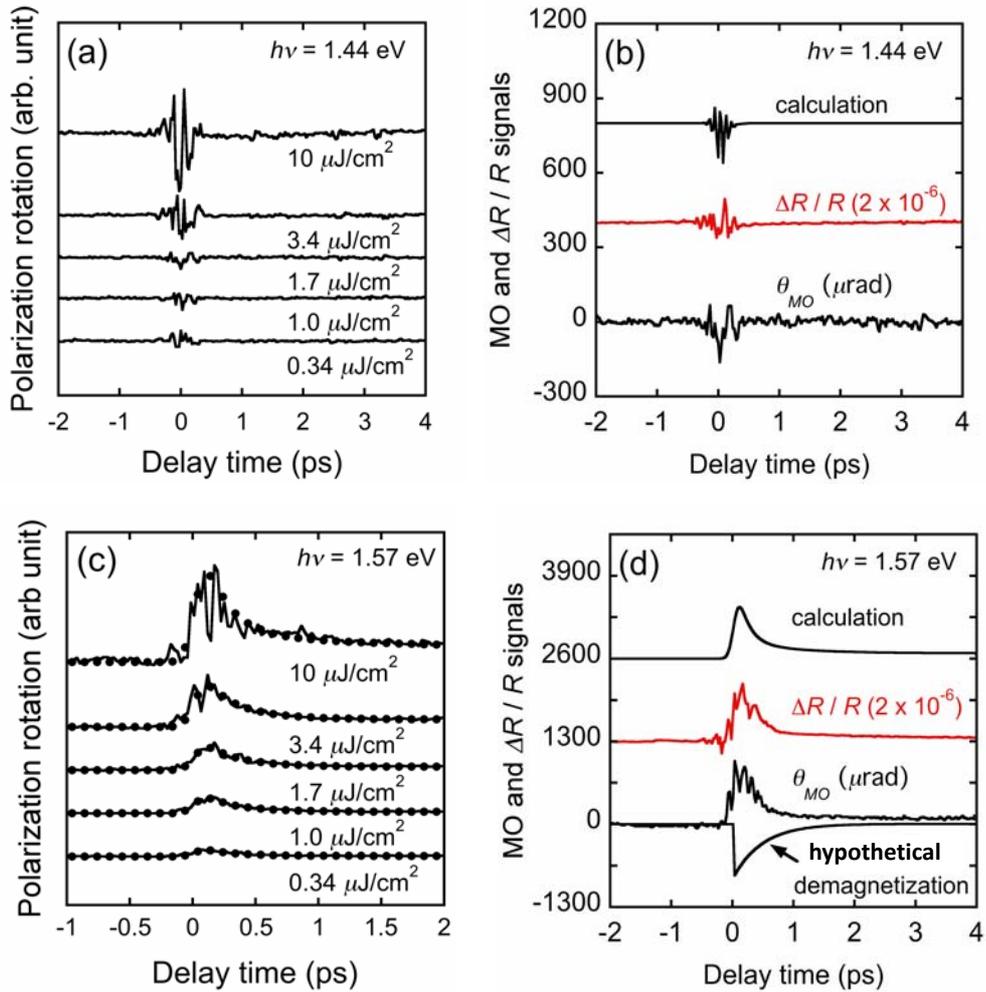

Figure 5: (a) Temporal MO profiles obtained from the $x = 0.02$ sample by below-gap excitation with five different pump fluences. (b) Three different profiles, from the top, calculation using Eqs. 1(a) and (b), experimental differential reflectance profile, and experimental polarization rotation profile. Both experimental data were obtained at the pump fluence $I_{pump} = 1.7$ μJ/cm². (c) Temporal MO profiles obtained from the same sample by above-gap excitation with five different pump fluences, together with calculation using Eq. 2 (dots). (d) Four different profiles, from the top, calculation with autocorrelation function using Eq. 2, experimental differential reflectance profile, experimental polarization rotation profile, and calculated ultrafast demagnetization profile. Both experimental data were obtained at the pump fluence $I_{\text{pump}} = 1.7\ \mu\text{J/cm}^2$. Profiles are intentionally shifted vertically for clarity.

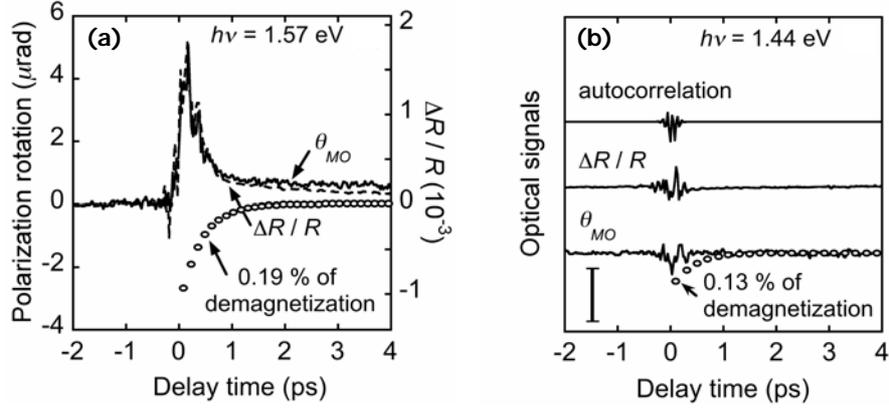

Figure 6: Two examples of calculated temporal MO profiles due to hypothetical, ultrafast demagnetization using Eq. 3. (a) the one (open circles) plotted together with the spike-like component obtained by measurements with P-P photon energy of $h\nu$ = 1.57 eV, and (b) the other (open circles) plotted together with the rapidly oscillating component obtained by measurements with P-P photon energy of $h\nu$ = 1.44 eV. The length of a vertical bar in the panel (b) represents polarization rotation of 2 μrad. For all experimental data, $I_{pump}$ = 1.7 μJ/cm$^2$ and $I_{probe}$ = 84 nJ/cm$^2$.

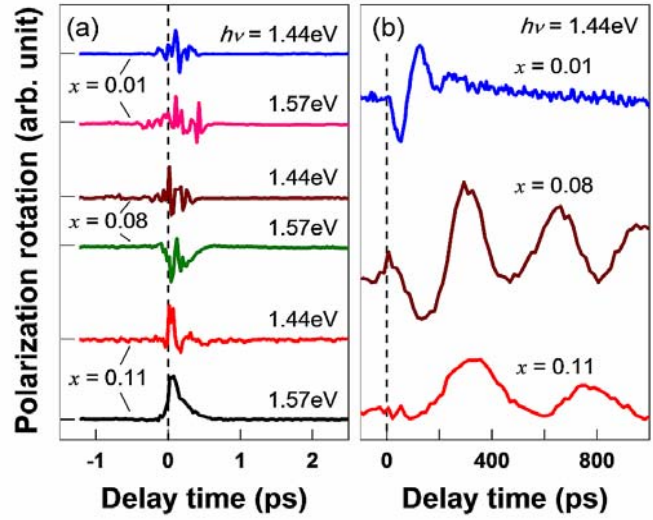

Figure 7: (a) MO profiles obtained from $x$ = 0.01, 0.08, and 0.11 samples with below- and above-gap excitations in (a) ps time sclae and (b) ns time scale. Pump and probe fluences are $I_{pump}$ = 1.7 μJ/cm$^2$ and $I_{probe}$ = 84 nJ/cm$^2$, respectively.

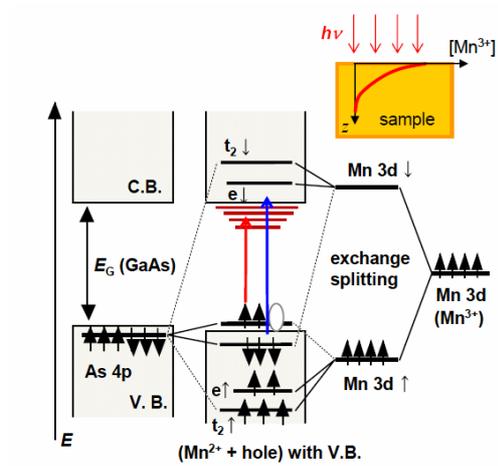

Figure 8: Schematic illustration of electronic structure of (Ga,Mn)As resulting from *p-d* hybridization between As 4*p* states in the GaAs valence band (VB) and Mn 3*d* states in $Mn^{2+}$ ions substituting for Ga sites. Joint Urbach tails are represented by horizontal lines (brown, online) below the edge of the conduction band (CB). Below- and above-gap excitations are represented by a short (red online) and a long (blue, online) arrows, respectively. Inset in the upper right depicts excitation-induced spatial gradient of photo-excited $Mn^{2+}$ ($Mn^{2+,*}$) ions.